\pdfoutput=1

\documentclass[11pt]{article}

\usepackage[final]{acl}

\usepackage{times}
\usepackage{latexsym}
\usepackage{amsmath}
\usepackage[T1]{fontenc}

\usepackage[utf8]{inputenc}

\usepackage{microtype}

\usepackage{inconsolata}

\usepackage[pdftex]{graphicx}
\usepackage{tabularx, booktabs}
\usepackage[most]{tcolorbox}
\title{Refining Text Generation for Realistic Conversational Recommendation \\
via Direct Preference Optimization}

\author{Manato Tajiri \hspace{0.7cm} Michimasa Inaba\\
  The University of Electro-Communications \\ 
  1-5-1, Chofugaoka, Chofu, Tokyo, Japan \\
  \texttt{t2530085@gl.cc.uec.ac.jp, m-inaba@uec.ac.jp} \\}

\begin{document}
\maketitle
\begin{abstract}
Conversational Recommender Systems (CRSs) aim to elicit user preferences via natural dialogue to provide suitable item recommendations. However, current CRSs often deviate from realistic human interactions by rapidly recommending items in brief sessions. This work addresses this gap by leveraging Large Language Models (LLMs) to generate dialogue summaries from dialogue history and item recommendation information from item description. This approach enables the extraction of both explicit user statements and implicit preferences inferred from the dialogue context. We introduce a method using Direct Preference Optimization (DPO) to ensure dialogue summary and item recommendation information are rich in information crucial for effective recommendations. Experiments on two public datasets validate our method's effectiveness in fostering more natural and realistic conversational recommendation processes. Our implementation is publicly available at: \url{https://github.com/UEC-InabaLab/Refining-LLM-Text}
\end{abstract}

\section{Introduction}
\begin{figure}[t!]
    \centering
    \includegraphics[width=\columnwidth]{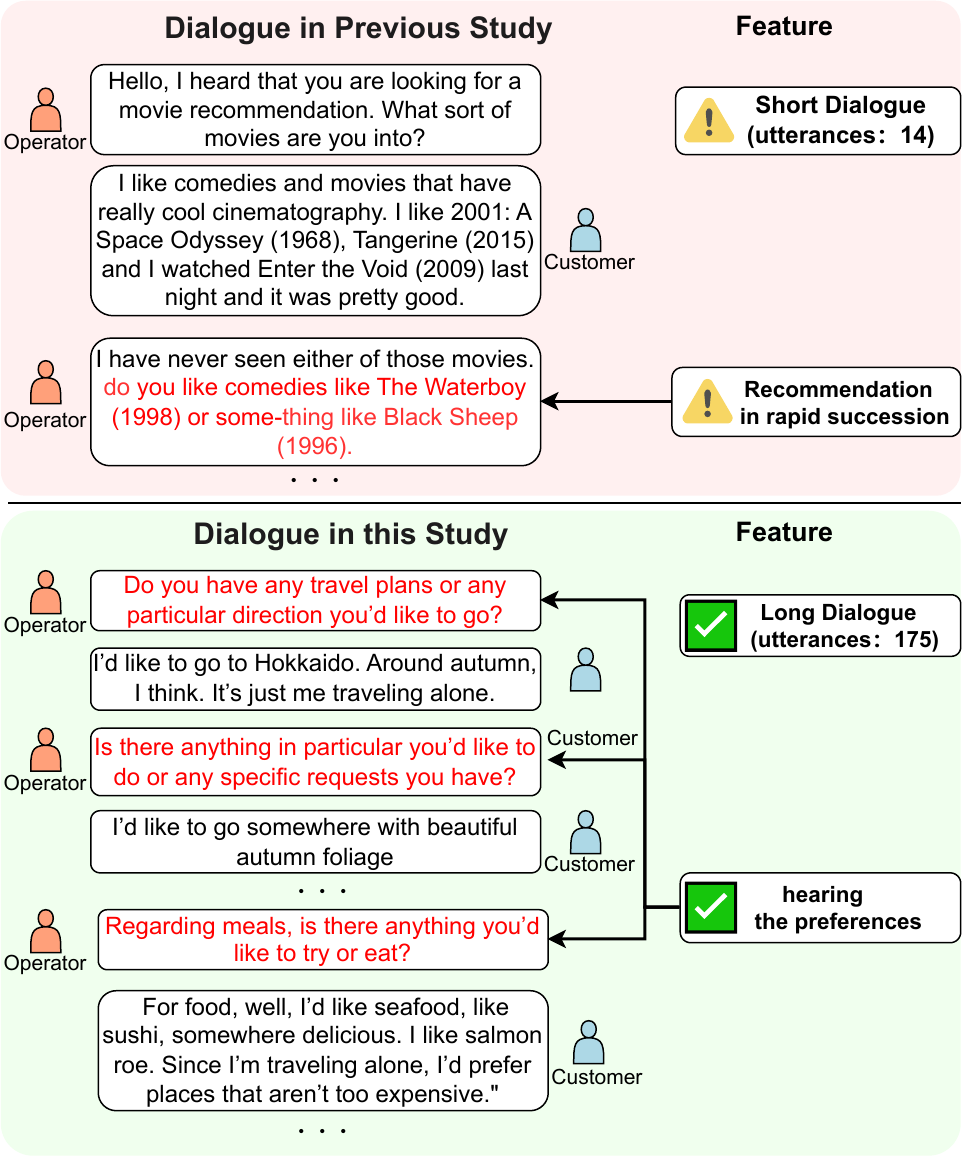}
    \caption{Comparison of recommendation dialogue corpus. The upper part shows a short movie recommendation dialogue example from REDIAL with rapid recommendation features. The lower part illustrates a tourist spot recommendation dialogue example from Tabidachi Corpus.}
    \label{dialogue}
\end{figure}
Recommender systems, pivotal for user satisfaction (e.g., from Amazon and Netflix ~\citep{amazon,netflix1,netflix2}), often face the cold-start problem with new users or items. Conversational Recommender Systems (CRSs) offer a promising solution by gathering user preferences through dialogue ~\citep{pre-CRS,crs-1,crs-2, chatcrs, Enpowering}. CRSs have advantages, mainly by gathering preferences via natural conversations. This approach obviates formal rating inputs, allowing information to be obtained naturally and improving accessibility; they can dynamically tailor inquiries to user responses; and they effectively elicit latent needs and interests users may be unaware of, particularly for new users through guided questioning.

However, many existing CRSs ~\citep{redial,tg-redial,crs-3,crs-4,crs-5,crs-6,crs-7} predominantly recommend items prematurely within brief dialogue sessions, often basing subsequent suggestions on immediate user feedback. This methodology starkly contrasts with realistic human conversational scenarios, where recommenders typically first elicit comprehensive information about a user's preferences, experiences, and context before suggesting carefully selected items, as illustrated by contrasting examples in Figure \ref{dialogue} (e.g., from REDIAL ~\citep{redial} and Tabidachi Corpus ~\citep{dataset}). Moreover, while implicit information—such as context, sentiment, and past experiences not explicitly articulated—plays a crucial role in natural interactions, its effective integration remains under-explored in current research. This discrepancy with natural dialogue processes is a key factor limiting the practical utility of existing systems. Our research, therefore, aims to foster more natural dialogue processes and enable the effective integration of such implicit information.

To address these aforementioned challenges, we turned our attention to the SumRec approach ~\citep{pre-research}, positing it as an effective strategy for tackling the ``divergence from natural dialogue processes'' and ``insufficient integration of implicit information'' inherent in conventional CRSs. SumRec leverages Large Language Models (LLMs) to generate dialogue summaries from dialogue history, aiming to extract both explicit and implicit user preferences. Concurrently, it generates item recommendation information from item descriptions to articulate item relevance to user preferences and experiences in natural language, rather than merely enumerating features. This dual-generation process facilitates a recommendation flow more akin to natural human-to-human dialogues, where appropriate items are recommended after a thorough understanding of the user's information. Despite its merits, a key limitation of SumRec is that its generated summaries or recommendation texts may sometimes lack information crucial for effective downstream recommendation tasks (e.g., item selection or scoring), potentially hindering the system's ability to interpret the relationship between user needs and item suitability. To overcome this, the LLM needs to be guided to extract and generate precisely the information essential for the recommendation process. Furthermore, SumRec's applicability was not extensively validated on general, realistic recommendation dialogues beyond specific domains like tourist recommendations from chit-chat. Therefore, this study proposes a recommendation method based on enhancing SumRec, applicable to more general and realistic conversational recommendation scenarios, aiming for improved recommendation quality.

The application of DPO ~\citep{dpo}, a method for fine-tuning LLMs based on preference data, aims to enable the generation of dialogue summaries and item recommendation information that are rich in content essential for item recommendation, thereby facilitating more accurate and appropriate recommendations.

The main contributions of this research are twofold.
(1) We propose an extension to SumRec by fine-tuning the LLM using DPO, creating a recommendation method tailored for realistic conversational recommendation datasets.
(2) We demonstrate through comparisons with baseline methods and the original SumRec that our proposed approach achieves superior recommendation performance on these datasets.

\section{Related work}

\subsection{Conversational Recommender System}
Existing conversational recommendation datasets ~\citep{pearl,redial,tg-redial,inspired,durecdial-2.0} predominantly feature scenarios where multiple items are recommended rapidly or in quick succession within short dialogue sessions, with subsequent recommendations determined by user feedback. Conversational Recommender Systems (CRSs) developed using these datasets ~\citep{PECRS,cr-walker,cola,crfr,cp-rec} are consequently tailored for such specific interaction patterns. This focus, however, limits their applicability and effectiveness in more realistic and nuanced conversational recommendation scenarios, where interactions might be longer, and recommendations are made more deliberately.

\subsection{Dialogue Summarization using LLMs}
Large Language Models (LLMs) like GPT~\citep{gpt}, Llama~\citep{llama}, and PaLM~\citep{palm} have demonstrated remarkable success across various AI research domains. Our work leverages LLMs to generate dialogue summaries from dialogue history, aiming to extract user preferences crucial for recommendations.
Several approaches have explored LLM-based dialogue summarization. \citet{dialogue-sum1} proposed generating factual summaries using smaller language models with GPT-3.5-Turbo as a teacher for contrastive learning. However, their method primarily targets short dialogues and faces challenges with longer conversations. \citet{dialogue-sum2} addressed long dialogues by pre-training Transformer-based models, though pre-training typically requires substantial data and computational resources. In contrast, our research focuses on achieving high-quality text generation through fine-tuning, without extensive pre-training. For longer dialogues, \citet{dialogue-sum3} proposed SummN, a method fine-tuning LLMs via supervised learning to first summarize dialogue chunks and then create a final summary from these. Our work adopts their method to generate dialogue summaries from dialogue history.

\subsection{Recommendation via Augmentation or Refinement of Item Description}
In our research, we facilitate item recommendation by generating item recommendation information that explain the suitability of an item for a particular user.
Similar to our work, \citet{llm-rec} proposed a method that recommends items by augmenting item description using LLMs. However, their study does not explicitly incorporate user preferences or experiences derived from the ongoing conversation into the augmentation of item description or the generation of item recommendation information.
\citet{ucepic} introduced a method to generate more appropriate item recommendation information by imposing vocabulary constraints during the generation process. Other approaches involve retrieving similar reviews or other external information using external tools and leveraging them to generate item recommendation information ~\citep{explainable,maple,xrec}.
However, these existing studies primarily focus on generating item recommendation information using historical user behavior data or past reviews. Their application is therefore challenging in scenarios like ours, where only the current dialogue history and item description are available. Consequently, our research proposes a method to generate pertinent item recommendation information based solely on user preferences and experiences derived from the dialogue history, in conjunction with item description.

\section{Method}
In this research, we extend SumRec to propose a method capable of high-performance recommendation on realistic conversational recommendation datasets.

\subsection{Task Definition}
This study focuses on item recommendation within conversational settings. Let $u_{o_i}$ denote an utterance from the operator (recommender) and $u_{c_i}$ denote an utterance from the customer (recommendee). The task addressed in this research is as follows. given a dialogue history $C=\{u_{o_1},u_{c_1},\ldots, u_{o_{n-1}},u_{c_{n-1}} \}$, a set of candidate items $T=\{t_1,\ldots , t_M\}$ at that point, and item description $D=\{d_1,\ldots , d_M\}$ for candidates, the objective is to predict the correct item $t_k$ that will be included in the next operator's utterance $u_{o_n}$.

\subsection{SumRec}

\begin{figure}[t!]
    \centering
    \includegraphics[width=0.5\textwidth]{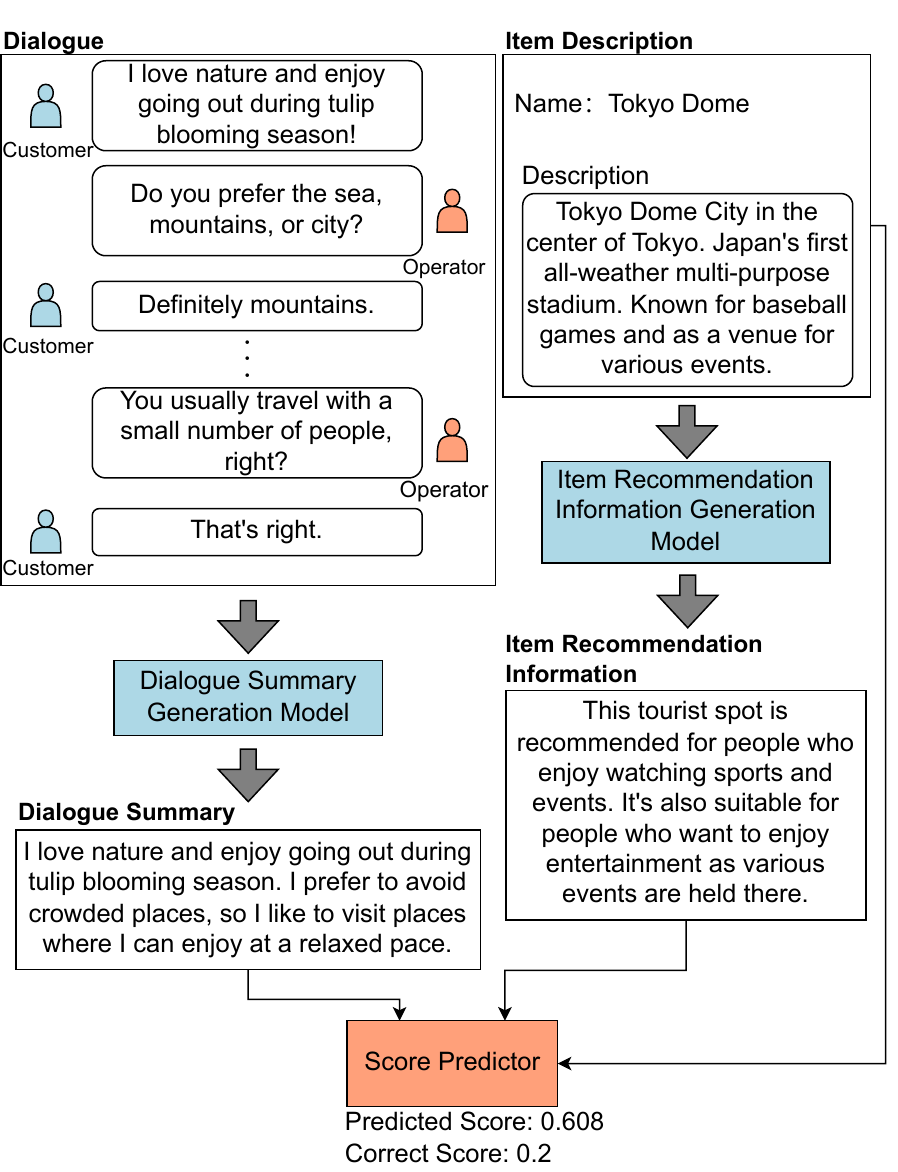}
    \caption{Item recommendation flow in SumRec. Dialogue Summaries and Item Recommendation Information, generated from Dialogue History and Item Descriptions respectively, are fed with the Item Description into a Score Predictor to estimate a recommendation score.}
    \label{pre}
\end{figure}

Figure \ref{pre} illustrates the item recommendation flow in SumRec. It inputs the dialogue summary, item recommendation information, and item description into a score predictor to predict a score, thereby recommending items. A limitation of SumRec is that the item recommendation information may not always include information about what kind of user an item is suitable for, which can lead to incorrect score predictions. While simple prompt engineering can instruct the LLM to, for instance, ``include user preference information,'' it is challenging to make the LLM selectively choose information crucial for the specific recommendation task. What is critical for our recommendation task is whether the generated texts (dialogue summary and item recommendation information) contain information that allows the score predictor to make accurate predictions. Therefore, a method that directly optimizes for this is required.
The following sections detail the recommendation procedure of SumRec.

\subsubsection{Dialogue Summary Generation Model}
The dialogue summary generation model uses an LLM to generate a summary from the dialogue history $C=\{u_{o_1},u_{c_1},\ldots, u_{o_{n-1}},u_{c_{n-1}} \}$. This module is essential for extracting information useful for recommendation from the dialogue history to perform more effective recommendations. The summarization process aims to ensure that the dialogue summary includes crucial information for recommendation, such as the speaker's preferences and experiences.
Experiments using Tabidachi Corpus, discussed later, involve datasets closely resembling actual dialogue scenarios, where dialogue histories are long, making it difficult to generate a summary in a single pass. Therefore, inspired by the method of Zhang et al. ~\citep{dialogue-sum3}, we first generate partial summaries by dividing the dialogue history into chunks and summarizing each chunk. Then, a final dialogue summary is generated based on these partial summaries. The actual prompts used are described in Appendix \ref{appendix:sum_prompt}.

\subsubsection{Item Recommendation Information Generation Model}
Item description primarily consists of objective facts, often lacking details about what kind of user would find the item recommendable. Therefore, SumRec employs an LLM to create item recommendation information based on the item description $D=\{d_1,\ldots , d_M\}$ of candidate items. The difference between item description and an item recommendation information is exemplified in Figure \ref{pre} by the inclusion of a phrase like ``It's also suitable for people who want to enjoy entertainment as various
events are held there.'' in the item recommendation information. The actual prompts used are detailed in Appendix \ref{appendix:rec_prompt}.

\subsubsection{Score Predictor}
The dialogue summary and item recommendation information obtained from the above processes, along with the item description, are concatenated using a [SEP] token and fed into a score predictor. This predictor estimates the recommendation score of an item for the customer. The model used for the score predictor is a pre-trained language model based on a Transformer encoder. It is trained as a regression task, where items recommended within the dialogue are assigned a target score of \(y=1\), and all other items are assigned \(y=0\). In the experiments described later, DeBERTa ~\citep{deberta} was used as the score predictor.

\subsection{Improving Information Extraction Performance using DPO}

\begin{figure*}[t!]
    \centering
    \includegraphics[width=\textwidth]{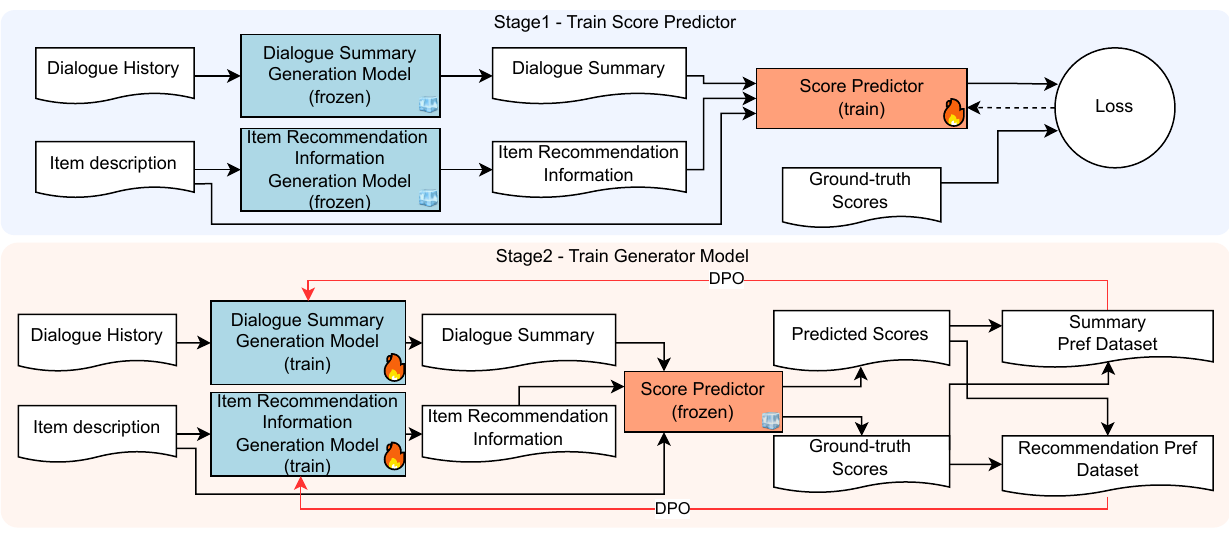}
    \caption{Process flow of the proposed method during training. The method employs a two-stage training procedure. In Stage 1, the Score Predictor is trained. In Stage 2, the Dialogue Summary Generation Model and the Item Recommendation Information Generation Model are trained using DPO.}
    \label{flow}
\end{figure*}

In this study, we employ Direct Preference Optimization (DPO) ~\citep{dpo} to generate texts (dialogue summaries and item recommendation information) that sufficiently contain information necessary for recommendation. While SumRec also aimed to include necessary information, relying solely on prompt engineering made it difficult to consistently extract and generate crucial details, often resulting in outputs that were abstract or overly generic.

Our research aims to enable appropriate recommendations by applying DPO to the LLM, thereby generating dialogue summaries and item recommendation information that adequately include information essential for item recommendation. Figure \ref{pre} shows a recommendation example from SumRec where there is a large discrepancy between the predicted score and the ground-truth score. In this example, the item recommendation information fails to include the information that ``a large number of people gather at Tokyo Dome'', leading to an incorrect score prediction for Tokyo Dome. Therefore, our work aims to generate texts that appropriately include information critical for recommendation.

Unlike SumRec, which does not fine-tune its dialogue summary generation model or item recommendation information generation model, our proposed method trains these models using DPO. This is to ensure that the score predictor can properly interpret the relationship between user preferences, experiences, and item description. The dataset for DPO training is created based on the prediction scores from the score predictor.

A key aspect of our approach is that the candidate texts are generated using well-designed, structured prompts. Consequently, the ``loser'' samples in our preference data are not nonsensical negative examples, but rather tend to be ``nearly good'' texts that are simply less effective for the recommendation task. This allows the DPO process to focus on learning fine-grained distinctions between high-quality and slightly less effective texts. An example of the preference data used for DPO training is provided in Appendix \ref{appendix:case-study}.

The training flow of our proposed method is depicted in Figure \ref{flow}. The training consists of two steps. Step 1 involves pre-training the score predictor, and Step 2 involves DPO-based training of the dialogue summary generation model and the item recommendation information generation model.

\subsubsection{Training the Score Predictor}
The preference data used for DPO training of the dialogue summary and item recommendation information generation models are created based on the output of the score predictor. Therefore, the score predictor is trained first. We use DeBERTa for the score predictor. The score prediction using DeBERTa can be expressed as Equation \ref{deberta}. The dialogue summary \( s \), item recommendation information \( r \), and item description \( d \) are concatenated with [SEP] tokens and input into the score predictor to obtain the predicted score $\hat{y}$.

\begin{equation}
\label{deberta}
\hat{y} = \text{DeBERTa}(s, r, d)
\end{equation}

The dialogue summary \( s \) is generated from the dialogue history using the dialogue summary generation model, and the item recommendation information \( r \) is generated from the item description using the item recommendation information generation model.

\subsubsection{Training the Dialogue Summary Generation Model}
This section describes the training procedure for the dialogue summary generation model. Let $\{s^n_1, \ldots, s^n_K\}$ be $K$ dialogue summaries, where each $s^n_k$ is a final summary. These are generated by an LLM from a given dialogue history $C_n$ by first creating a set of $M$ partial summaries $\{ps^n_1, \ldots, ps^n_M\}$ from $C_n$, concatenating them into a combined text defined as $PS_n$, and then generating $s^n_k$ from this $PS_n$. Let \(\{r^n_1,\ldots, r^n_{M_n} \}\) be the item recommendation information generated by an LLM for the candidate item descriptions \(\{d^n_1,\ldots, d^n_{M_n} \}\). The score prediction is given by Equation \ref{eq:score_prediction}.

\begin{equation}
\label{eq:score_prediction}
\hat{y}^n_{k,m} = \text{DeBERTa}(s^n_k, r^n_m, d^n_m)
\end{equation}

For each dialogue summary, item recommendation information \(r^n_m\), and item description \( d^n_m \) input into the score predictor, the absolute difference between the output score \(\hat{y}^n_{k,m}\) and the ground-truth score \( y^n_m \) is calculated. The dialogue summary closest to the ground-truth score and the one furthest from it are determined as shown in Equations \ref{eq:best_summary} and \ref{eq:worst_summary}, respectively.

\begin{equation}
\label{eq:best_summary}
s^n_{m,+} = \arg\min_{s^n_k} |y^n_m - \hat{y}^n_{k,m}|
\end{equation}

\begin{equation}
\label{eq:worst_summary}
s^n_{m,-} = \arg\max_{s^n_k} |y^n_m - \hat{y}^n_{k,m}|
\end{equation}

These pairs are used as preference data for DPO. The DPO loss function is then expressed as in Equation \ref{eq:DPO_loss}.

\begin{align}
\mathcal{L}_{\text{DPO}} = 
&-\mathbf{E}_{(PS_n, {s}_{m,\text{+}}^{n}, {s}_{m,\text{-}}^{n}) \sim \{m \in \mathcal{M}_n, n \in \mathcal{N}\} } \notag \\
&\Bigg[ \log \sigma \Bigg( 
\beta \log \frac{\pi_\phi({s}_{m,\text{+}}^{n}|PS_n)}{\pi_{\phi_{\text{ref}}}({s}_{m,\text{+}}^{n}|PS_n)} \notag \\
&\quad - \beta \log \frac{\pi_\phi({s}_{m,\text{-}}^{n}|PS_n)}{\pi_{\phi_{\text{ref}}}({s}_{m,\text{-}}^{n}|PS_n)} 
\Bigg) \Bigg] \label{eq:DPO_loss}
\end{align}

Here,$\mathcal{N}$ is the set of indices for dialogue histories $\{ 1, 2, \ldots , |N| \}$, $\mathcal{M}_n$ is the set of indices for candidate items for dialogue history $C_n$($n \in \mathcal{N}$)  i.e., $\{ 1, 2, \ldots , |M_n| \}$, \(\beta\) is the temperature parameter, \(\pi_{\phi_{\text{ref}}}\) is the output probability of the dialogue summary generation model before training, \(\pi_{\phi}\) is the output probability of the dialogue summary generation model being trained, and \(\sigma\) is the sigmoid function.The hyperparameters used in the experiments are detailed in Appendix \ref{appendix:hyper-parameter}.

\subsubsection{Training the Item Recommendation Information Generation Model}
The item recommendation information generation model is also trained using DPO in a similar manner. This aims to enable the model to generate item recommendation information that better incorporate information crucial for item recommendation. From the candidate item descriptions \(\{d^n_1,\ldots, d^n_{M_n} \}\), let $d^n_m$ be the item description for which the ground-truth score is 1. Let \(\{r^n_{m,1},\ldots, r^n_{m,J} \}\) be $J$ item recommendation information generated by an LLM based on $d^n_m$. The reason for not using item description with a ground-truth score of 0 is to avoid potentially training the model to consider sentences lacking necessary recommendation information as good outputs. Let $s^n$ be the dialogue summary generated by an LLM from a given dialogue history \(C_n\). Each item recommendation information, along with item description $d^n_m$ and dialogue summary $s^n$, is input into the score predictor. The absolute difference between the output score and the ground-truth score \( y^n_m \) is calculated.
The item recommendation information closest to the ground-truth score is denoted as \({r}_{m,\text{+}}^{n}\), and the one furthest is denoted as \({r}_{m,\text{-}}^{n}\). These are used as preference data. The loss function for training the item recommendation information generation model is analogous to Equation \ref{eq:DPO_loss}, where the policy generates item recommendation information  $r$ conditioned on item description $d^n_m$ instead of a summary $s$ conditioned on dialogue history \(PS_n\).

\section{Experiment}
To evaluate our proposed method, we conducted a comparative analysis against baselines through automatic evaluation. The LLM used as the text generation model (for both dialogue summaries and item recommendation information) in this study was Llama-3.1-Swallow-8B-v0.1 ~\citep{Okazaki:COLM2024,Fujii:COLM2024}\footnote{https://huggingface.co/tokyotech-llm/Llama-3.1-Swallow-8B-v0.1}, and the model used for the score predictor was deberta-v3-japanese-large (352M parameters)~\citep{deberta}\footnote{https://huggingface.co/globis-university/deberta-v3-japanese-large}.

\subsection{Datasets}
Experiments were conducted using two Japanese datasets: Tabidachi travel agency task dialogue corpus~\citep{dataset} and ChatRec~\citep{pre-research}.

Tabidachi Corpus features tourist spot recommendation dialogues between an operator and a customer planning a sightseeing trip via Zoom. The operator uses a system to find tourist information while conversing, and the customer decides on a travel plan based on a predefined scenario. Further details are in Appendix \ref{appendix:tabidachi_corpus}.

ChatRec, though not representing realistic recommendation dialogues, was used as SumRec's evaluation dataset and is thus included here. It comprises chit-chat dialogues between two CrowdWorks participants (minimum 10 turns each) under a ``strangers in a waiting room'' scenario. Data was collected under three topic conditions: Travel, Except for Travel, and No Restriction. Details are in Appendix \ref{appendix:chatrec}.

\subsection{Evaluation Metrics}
Since our task involves selecting an item to recommend from a set of candidates, we evaluated the recommendation performance using Hit Rate (HR) and Mean Reciprocal Rank (MRR), which are common metrics for retrieval tasks.

\subsection{Baseline and Implementation}
To demonstrate the effectiveness of our proposed method, we conducted experiments and evaluations against the following two baselines. The hyperparameters used during implementation are detailed in Appendix \ref{appendix:hyper-parameter}.

\begin{itemize}
    \item \textbf{Baseline}:The dialogue summary is generated using the LLM (Llama-3.1-Swallow-8B-v0.1) without DPO. Only this dialogue summary and the item description are input to the score predictor. No item recommendation information is generated or utilized by this model.
    \item \textbf{SumRec}:This model uses the LLM (Llama-3.1-Swallow-8B-v0.1) without DPO to generate both the dialogue summary and the item recommendation information. The generated dialogue summary, item recommendation information, and item description are then fed into the score predictor.
\end{itemize}

\subsection{Results}
Table \ref{result} shows the comparison results of Hit Rate (HR) and Mean Reciprocal Rank (MRR) for both Tabidachi Corpus and ChatRec datasets.
On Tabidachi Corpus, our proposed method outperformed existing methods across all rank cutoffs, with particularly significant performance improvements observed at higher ranks. This implies that our method can substantially enhance the quality of the candidate list that users typically view first.

Although ChatRec inherently presents a recommendation task with a high baseline performance, our proposed method maintained HR levels comparable to or exceeding existing methods, while consistently achieving the best MRR. This result suggests that our approach can stably improve the precision at early ranks even in situations with different dialogue densities and domains.
Consequently, these findings confirm that our proposed method improves the quality of top-position recommendations across diverse datasets and can contribute to the rapid and highly accurate recommendations required in practical applications.

\begin{table}[t]
\centering
\resizebox{\columnwidth}{!}{
\begin{tabular}{lllccc}
\hline
Dataset & Method & Metrics & @1 & @3 & @5 \\ \hline
Tabidachi Corpus & Baseline  & HR $\uparrow$  & 0.2439 & 0.5056 & 0.7146 \\
          &           & MRR $\uparrow$ & 0.2439 & 0.3587 & 0.4057 \\ \cline{2-6}
          & SumRec    & HR $\uparrow$  & 0.2040 & 0.5376 & \textbf{0.7574} \\
          &           & MRR $\uparrow$ & 0.2040 & 0.3527 & 0.4032 \\ \cline{2-6}
          & Ours      & HR $\uparrow$  & \textbf{0.2474} & \textbf{0.5525} & 0.7231 \\
          &           & MRR $\uparrow$ & \textbf{0.2474} & \textbf{0.3796} & \textbf{0.4181} \\ \hline
ChatRec  & Baseline  & HR $\uparrow$  & 0.8423 & 0.9799 & 0.9933 \\
         &           & MRR $\uparrow$ & 0.8423 & 0.9049 & 0.9081 \\ \cline{2-6}
         & SumRec    & HR $\uparrow$  & 0.8255 & 0.9698 & \textbf{1.0} \\
         &           & MRR $\uparrow$ & 0.8255 & 0.8915 & 0.8984 \\ \cline{2-6}
         & Ours      & HR $\uparrow$  & \textbf{0.8591} & \textbf{0.9832} & 0.9933 \\
         &           & MRR $\uparrow$ & \textbf{0.8591} & \textbf{0.9172} & \textbf{0.9196} \\ \hline
\end{tabular}}
\caption{Proposed Methodology and Baseline Evaluation Results}
\label{result}
\end{table}

\begin{table}[t!]
\centering
\small
\begin{tabularx}{\columnwidth}{@{} X c c c c @{}} 
\hline
Method & Avg. Len. & Distinct-1/2 & BLEU  & ROUGE-L \\ 
\hline
\multicolumn{5}{@{}l}{\textbf{Dialogue Summary}} \\ 
SumRec   & 118.6 & 0.251 / 0.611  & -- & -- \\
Proposed & 151.2 & 0.187 / 0.526  & -- & -- \\
\hline
\multicolumn{5}{@{}l}{\textbf{Item Recommendation Information}} \\ 
SumRec   & 149.7 & 0.247 / 0.586 & 3.608  & 0.087 \\
Proposed & 247.2 & 0.164 / 0.433 & 1.455 & 0.019 \\
\hline
\end{tabularx}
\caption{Automatic analysis of dialogue summaries and item recommendation information. For the item recommendation information, BLEU and ROUGE-L were calculated against the original item descriptions as references.}
\label{tab:auto-metrics} 
\end{table}

\subsection{Analysis of Generated Texts}
The quantitative analysis using automatic metrics, presented in Table \ref{tab:auto-metrics}, indicates that the application of DPO in our proposed method induced notable changes in the structure and lexical usage of the generated texts.Example of generated text is in Appendix \ref{appendix:case-study}.

Firstly, for dialogue summaries, the proposed method produced significantly longer sentences than SumRec. This suggests an enhanced capability to retain more detailed user preferences and conversational context. Conversely, the Distinct-1 and Distinct-2 scores decreased, indicating an increased tendency for the model to repeatedly use the same keywords. This phenomenon can be interpreted as a consequence of DPO guiding the model to prioritize and preserve phrases deemed important by the score predictor.

Regarding item recommendation information, a similar trend of increased length and reduced lexical diversity was concurrently observed. Furthermore, n-gram similarity metrics such as BLEU and ROUGE-L, calculated against the original item descriptions as references, also exhibited a decrease. This suggests that the proposed method places greater emphasis on incorporating explanatory elements beneficial for recommendation, rather than prioritizing superficial n-gram overlap with these source item descriptions. These observations imply that both dialogue summaries and item recommendation information were optimized towards ``adequately containing information necessary for the task.''

\subsection{Ablation Study}

\begin{table}[t!]
\centering
\small
\resizebox{\columnwidth}{!}{
\begin{tabular}{llccc}
\hline
Method & Metrics & @1 & @3 & @5 \\ \hline
Ours     & HR $\uparrow$  & \textbf{0.2474} & 0.5525 & 0.7231 \\
         & MRR $\uparrow$ & \textbf{0.2474} & \textbf{0.3796} & 0.4181 \\ \cline{1-5}
w/o Rec-DPO & HR $\uparrow$  & 0.2393 & \textbf{0.5560} & \textbf{0.7402} \\
         & MRR $\uparrow$ & 0.2393 & 0.3772 & \textbf{0.4195} \\ \cline{1-5}
w/o Sum-DPO & HR $\uparrow$  & 0.2341 & 0.5176 & 0.7363 \\
         & MRR $\uparrow$ & 0.2341 & 0.3554 & 0.4051 \\ \hline
\end{tabular}}
\caption{Ablation study on Tabidachi Corpus. ``w/o Rec-DPO'' ablates DPO from item recommendation information generation (i.e., DPO only on summary), while ``w/o Sum-DPO'' ablates DPO from dialogue summary generation (i.e., DPO only on recommendation information).}
\label{tab:ablation}
\end{table}

In Table \ref{tab:ablation}, we individually evaluated a method that optimizes only the dialogue summary generation model with DPO (w/o Rec-DPO) and a method that optimizes only the item recommendation information generation model with DPO (w/o Sum-DPO).

The ``w/o Rec-DPO'' method surpassed SumRec on all metrics except for HR@5, with notable gains in HR and MRR, especially at higher ranks.This indicates that enhancing the quality of the summary allows user preferences to be reflected more precisely, significantly improving the relevance of initially presented items.
While ``w/o Sum-DPO'' also showed some improvement, its effect was not as pronounced as that of ``w/o Rec-DPO,'' and the performance gap tended to widen at higher ranks. This result suggests that while improving recommendation information offers a supplementary benefit, refining the dialogue summary, which forms the foundation of the recommendation process, is more critical. In contrast, the effect of DPO on the item recommendation information themselves was limited, which we will analyze in the next section.

Ultimately, ``Ours,'' which involves DPO training for both models, demonstrated the highest performance across all metrics. This confirms that by fine-tuning both the summary and recommendation information generation, the quality of user preference representation and item description is synergistically enhanced, further boosting recommendation accuracy.

\subsection{Human Evaluation}

\begin{figure}[t!]
    \centering
    \includegraphics[width=\columnwidth]{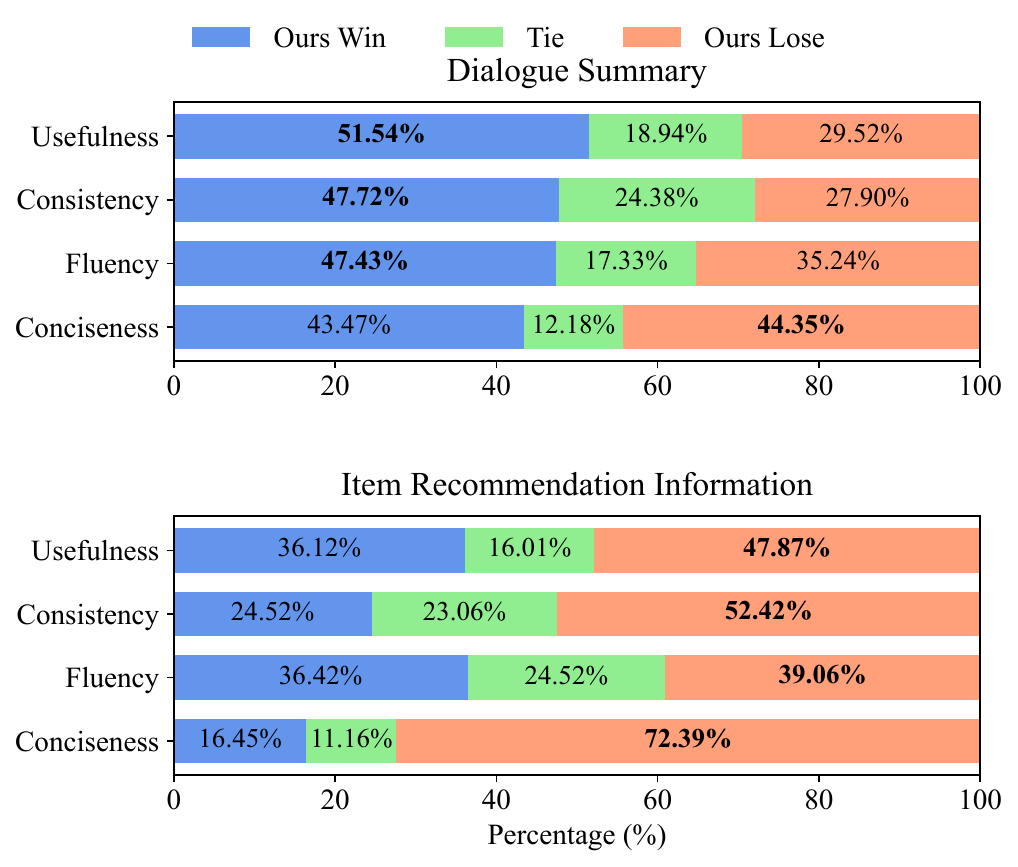}
    \caption{Human evaluation of the proposed method and SumRec on Tabidachi Corpus, assessing dialogue summaries and item recommendation information.}
    \label{human-eval}
\end{figure}

To evaluate generated dialogue summaries and item recommendation information, we conducted a human evaluation using CrowdWorks. Outputs from our proposed method and SumRec were compared on four criteria: Consistency, Conciseness, Fluency, and Usefulness. For 54 recommendation dialogues and their item descriptions, 10 CrowdWorkers evaluated each; Figure \ref{human-eval} illustrates these results.

Regarding dialogue summaries, our proposed method outperformed SumRec on Consistency, Fluency, and Usefulness. Approximately half the evaluators rated our summaries as superior, with ``Usefulness'' showing the most significant difference. This suggests DPO enhanced summary quality, enabling more accurate user preference capture. While ``Conciseness'' showed no substantial difference, our approach improved other aspects without compromising it, despite a slight tendency towards verbosity. For item recommendation information, however, SumRec performed better across all metrics. This finding may be related to the results of the ablation study, where applying DPO solely to item information did not yield consistent performance gains. However, this decline in quality is not considered a critical issue, as the item recommendation information is intended for internal use and is not directly presented to the user.

DPO training significantly improves dialogue summary quality. Our ablation study, detailed in Table \ref{tab:ablation}, showed the ``w/o Rec-DPO'' condition, using DPO-trained summaries, boosted recommendation performance. This suggests that enhanced dialogue summaries, particularly their improved extraction of recommendation-relevant information, are key drivers of overall system performance. Human evaluation further corroborated these findings, underscoring the critical role of DPO-trained dialogue summaries in enhancing recommendation system efficacy, as shown in Table \ref{tab:ablation}.

The above findings from the human evaluation corroborate that DPO training of the dialogue summary is particularly crucial for improving the recommendation system's performance, aligning with the results of the ablation study.

\section{Conclusion}

In this study, we proposed a method that optimizes both dialogue summary and item recommendation information generation models using Direct Preference Optimization (DPO), aiming for a system better suited to realistic conversational recommendation. Experimental results showed our method achieved superior recommendation performance over baselines on both datasets. DPO training for dialogue summaries significantly contributed to this, with human evaluation confirming enhanced extraction of recommendation-useful information. Future work includes further improving recommendation performance while maintaining the quality of generated item recommendation information. Additionally, addressing potential risks such as data-specific biases, content hallucination, and misuse will be crucial.

\section*{Limitations}
First, the models employed, Llama-3.1-Swallow-8B-v0.1 and DeBERTa-v3-japanese-large, are medium-scale and not representative of state-of-the-art large language models (LLMs) with hundreds of billions of parameters. While larger models could potentially enhance performance, they would also significantly increase GPU memory consumption and inference latency, creating a trade-off with operational costs that remains a challenge.

Second, a significant limitation is the narrow scope of our evaluation. The experiments were conducted exclusively on two Japanese datasets within the travel domain (Tabidachi Corpus and ChatRec). Therefore, the generalizability of our method to other domains and languages is yet to be verified.

Finally, a notable limitation is the persistence of hallucinations in the generation of item recommendation information and dialogue summaries, where the model fabricates features not present in the source content. Even when such fabrications are not directly presented to the user, they can adversely affect the model's explainability.

\bibliography{custom}

\appendix

\section{Datasets Details}
\label{appendix:datasets}
The statistics of the dataset used in this study are summarized in Table \ref{tab:dialogue_datasets}

\begin{table*}[htbp]
\centering
\setlength{\tabcolsep}{3pt} 
\begin{tabular*}{\textwidth}{@{\extracolsep{\fill}}c|c|cccc@{}} 
\hline
\multicolumn{1}{c|}{\bf Metric} & \multicolumn{1}{c|}{\bf Tabidachi Corpus} & \multicolumn{4}{c}{\bf ChatRec} \\
& & {\bf T} & {\bf E} & {\bf N} & {\bf ALL} \\
\hline
Dialogues & 165 & 237 & 223 & 545 & 1,005 \\
\multicolumn{1}{c|}{\it \scriptsize (Train / Val / Test)} & \it \scriptsize 126 / 15 / 24 & \it \scriptsize 189 / 13 / 35 & \it \scriptsize 178 / 12 / 33 & \it \scriptsize 436 / 28 / 81 & \it \scriptsize 803 / 53 / 149 \\ 
Utterances & 42,663 & 5,238 & 5,009 & 11,735 & 21,982 \\
\hline
\end{tabular*}
\caption{Statistics of Dialogue Datasets, including training, validation, and test splits for the number of dialogues.}
\label{tab:dialogue_datasets}
\end{table*}

\subsection{Tabidachi Corpus}
\label{appendix:tabidachi_corpus}
Table \ref{conversation} presents an example of a dialogue history included in the dataset. An example of information for a tourist destination that can actually be viewed by the tourist information retrieval system is shown in Table \ref{tab:sight_info}. In this study, we used the concatenation of ``Summary'' and ``Feature'' from Table \ref{tab:sight_info} as the item description.

For Tabidachi Corpus, a total of 55 participants acted as customers (recommendees): 25 adults, 10 elderly individuals, and 20 children. Each participant engaged in a total of six recommendation dialogues. For each customer participant, there are three dialogue datasets where the recommendation dialogue was conducted while sharing the screen of the tourist information retrieval system, and three dialogue datasets where the dialogue was conducted without screen sharing. In this research, we used only the dialogues conducted without screen sharing. This decision was made because the visual information received by the customer participant through screen sharing cannot be directly input into the LLM.

\begin{table}[ht]
  \centering
  \small
  \begin{tabularx}{\columnwidth}{@{}l X@{}}
    \hline
    Operator & Hello. Thank you for using our service today. Um, regarding your travel plans, um, do you have any particular destination in mind that you would like to visit?\\
    Customer & Yes. Um, I would like to go to Hokkaido.\\
    Operator & Ah, yes. Um, do you have any preference for the season?\\
    Customer & Um, around autumn, I think.\\
    Operator & Um, how many people are planning to go?\\
    Customer & Ah, just me, just myself alone.\\
    Operator & Ah, understood. \texttt{<>} I will look into it, so please wait a moment.\\
    Customer & Yes. Ah, yes. Yes, please do.\\
    Operator & Um, is there anything specific you'd like to do, or any particular preferences?\\
    Customer & Yes. Ah, well. Um, I'd like to go somewhere with beautiful autumn leaves.\\
    \ldots & \ldots \\
    Operator & Yes. Ah, there is one thing but...\\
    Customer & Yes.\\
    Operator & Um, \texttt{<>}, around Sapporo and Mount Hakodate, particularly, are there any other places you'd like to visit?\\
    Customer & Ah, yes. Around that area, if there are any recommendations.\\
    Operator & Let me see... also\\
    Customer & Yes.\\
    Operator & \texttt{<>}, it's near Sapporo but...\\
    Customer & Yes.\\
    Operator & There is a place called Satellite Place\\
    Customer & Yes.\\
    \hline
  \end{tabularx}
  \caption{Dialogue example from Tabidachi Corpus. (English Version, translated from the Japanese by the author)}
  \label{conversation}
\end{table}

\begin{table}[ht]
  \centering
  \small
  \begin{tabularx}{\columnwidth}{@{}p{0.12\columnwidth}|p{0.17\columnwidth}|X@{}}
    \hline
    SightID &        & 80042498 \\ \hline
    Title   &        & Former Sougenji Stone Gate (Kyuu Sougenji Ishimon) \\ \hline
    Detail  & Area   & Kyushu/Okinawa\texttt{>}Okinawa Prefecture\texttt{>}Naha/Southern Main Island \\ \cline{2-3}
            & Genre1 & See\texttt{>}Buildings/Historic Sites\texttt{>}Historical Structures \\ \cline{2-3}
            & Genre2 &  \\ \cline{2-3}
            & Summary & A triple-arch gate made of Ryukyu limestone. The massive stone gate extending nearly 100m was built using cut stone masonry technique and is designated as a National Important Cultural Property. The interior was the temple grounds where Sougenji Temple, which enshrined the spirits of the Sho Dynasty, once stood, but was completely destroyed during the Battle of Okinawa.\\ \cline{2-3}
            & Time &  \\ \cline{2-3}
            & Closed &  \\ \cline{2-3}
            & Price & Free to visit \\ \cline{2-3}
            & Tel & 098-868-4887 \\ \cline{2-3}
            & Address & 1-9-1 Tomari, Naha City, Okinawa Prefecture \\ \cline{2-3}
            & Station & Miebashi \\ \cline{2-3}
            & Parking & None \\ \cline{2-3}
            & Traffic1 & 10-minute walk from Yui Rail (Okinawa Monorail) Miebashi Station or Makishi Station \\ \cline{2-3}
            & Traffic2 & 6 km from Okinawa Naha Airport \\ \cline{2-3}
            & Feature & Takes about 30 minutes to visit / Recommended for women / Recommended for history enthusiasts \\ \cline{2-3}
            & Treasure & Important Cultural Property (Structure) \\ \hline
  \end{tabularx}
  \caption{Example of item description in Tabidachi Corpus. (English Version, translated from the Japanese by the author)}
  \label{tab:sight_info}
\end{table}

\subsection{ChatRec Dataset}
\label{appendix:chatrec}

The tourist destination information for recommendation consists of 3,290 domestic spots. These were obtained by starting with approximately 45,000 domestic spots listed on Rurubu and then excluding those with fewer than 100 TripAdvisor reviews. This collection of 3,290 spots was further organized into 147 files, with each file containing 10–20 spots grouped by prefecture. After each dialogue, one of these files was randomly assigned to the workers, who were then asked to rate the spots within that file (an average of 15.7 spots per file). Additionally, a ``human-predicted score,'' calculated as the average of interest scores estimated by five third-party workers, is provided for each spot. The prediction scores are on a 5-point scale from 1 to 5. In this study, we converted scores of 2 or less to ``dislike'' (0) and scores of 3 or more to ``like'' (1). We conducted experiments using this scoring method, which aligns with that of Tabidachi Corpus.Examples of dialogues and item description from ChatRec are presented in Table \ref{tab:chatrec-conversation} and Table \ref{tab:chatrec_rec-info}, respectively.

\begin{table}[ht]
\centering
\small
\begin{tabular}{@{}l p{0.82\columnwidth}@{}}
\hline
A & What are your plans for dinner?\\
B & Thank you in advance. I'm planning to make ginger pork for dinner today. How about you?\\
A & Since it's cold, I'm thinking about having shabu-shabu, but ginger pork sounds good too.\\
B & That sounds nice. But I've already prepared for ginger pork today, so I'm thinking about having shabu-shabu tomorrow.\\
A & Pork for two days in a row, do you like pork?\\
B & I do like pork. I prefer chicken or pork over beef.\\
A & Do you use pork for curry?\\
B & Yes. We usually make it with pork at home. Are you perhaps a beef person?\\
A & We use beef at home. Does that mean you live in the eastern region?\\
B & Not necessarily, but for some reason we've always used pork at my home.\\
A & I see, what's your favorite pork dish?\\
B & For pork dishes, I like wrapping cheese with pork and seasoning it with a sweet and savory sauce.\\
A & That's quite elaborate. Do you put only cheese inside?\\
B & Not at all. I also add shiso leaves.\\
A & Is this fried, or do you just grill it?\\
B & It's delicious when fried too, but I'm concerned about the calories, so currently I just grill it.\\
A & What's the best side dish for it?\\
B & I'm not sure if it's the best, but I usually serve it with lettuce and cherry tomatoes.\\
A & Just imagining it makes me hungry.\\
B & Indeed. Do you like beef?\\
A & I do! I love steak and yakiniku (grilled meat).\\
A & Thank you for your time!\\
\hline
\end{tabular}
\caption{Dialogue example from the ChatRec dataset. (English Version, translated from the Japanese by the author)}
\label{tab:chatrec-conversation}
\end{table}

\begin{table}[ht]
\small                      
\setlength{\tabcolsep}{3pt} 
\begin{tabular}{p{0.22\columnwidth}|p{0.78\columnwidth}}
\hline
id          &                7\\ \hline
name        &                Sumida Park\\ \hline
description &                Located alongside the Sumida River, it has long been known as a famous cherry blossom viewing spot. In spring, when approximately 500 cherry trees planted along the Sumida embankment bloom, the park becomes crowded with many flower-viewing visitors. The park, which extends from Azuma Bridge, features walking paths that make for an ideal strolling course. From the X-shaped Sakura Bridge, visitors can enjoy a view of the Sumida River below.\\ \hline
\end{tabular}
\caption{Example of item description in the ChatRec dataset. (English Version, translated from the Japanese by the author)}
\label{tab:chatrec_rec-info}
\end{table}

\section{Prompt}
\label{appendix:prompt}
The prompts used in this study are detailed below.
For Tabidachi Corpus, dialogues were segmented into chunks of 30 utterances each to generate partial summaries. Subsequently, these partial summaries were concatenated and then used to create the final comprehensive summary.
\subsection{Dialogue Summary Prompt}
\label{appendix:sum_prompt}

\newpage
\onecolumn

\begin{tcolorbox}[
  colback=gray!15,
  colframe=gray!70,
  coltitle=white,
  colbacktitle=gray,
  title=\texttt{Prompt for Generating Partial Summaries in Tabidachi Corpus}, 
  fonttitle=\bfseries\sffamily,
  fontupper=\footnotesize,
  boxrule=0.4mm,
  arc=2mm,
  left=2mm,
  right=2mm,
  top=1mm,
  bottom=1mm,
  sharp corners=south,
  enhanced jigsaw,
  breakable,
]

{\textbf{[TASK]}} \\ 
Summarize B's hobbies, preferences, habits, and profile regarding tourist destinations based on the dialogue history between A and B.\\
Please strictly adhere to the following points:\\
- Extract the conditions B seeks in a tourist destination.\\
- The output must be a single sentence, without line breaks.\\
- Do not structure the output.\\
- Do not include information not mentioned in the dialogue.\\
- Do not output URLs, etc.\\

\medskip
{\textbf{--Example1--}}\\\leavevmode
[Dialogue History]\\ 
A: First off, do you have any hobbies? \\
B: I'm an indoor person, so I mostly just play games at home. What about your hobbies? \\
A: I'm also an indoor type, so my hobbies are all things I can do at home, like watching movies or baking. What kind of games do you like? \\
B: I play challenging action games. Baking sounds nice. I also go for walks quite often. \\
A: Do you happen to have a dog? \\
B: I'd love to have a dog; it's a wish of mine. When I say walks, I don't like walking a lot, just around the neighborhood and that's it. \\
A: That's nice. I want to make walking a hobby too. I used to go for walks often when I had a dog before, but I don't know how to enjoy walking alone. \\
B: It's true that it's hard to set goals for walks. I just feel down if I stay home all the time, so I want to get some fresh air. \\
A: I can understand being excited if it's an unfamiliar place, but do you have any tips for enjoying walks in a familiar neighborhood? \\
B: It's true that new places are fun just to look at. It's a bit plain, but I also try to improve my health by being conscious of my posture when I walk. \\
A: Maybe it's good to be conscious of it being for health. I'm not good at running, so I at least want to move my body by walking. \\
B: There's a cherry tree nearby, so it's fun to walk there in spring. \\
A: It's rewarding if there's a destination or if the scenery along the way is good. Recently, there are also walking apps that can link with GPS, right? \\
B: Having a destination is great. I actually use a walking app. By the way, what kind of scenery do you like? \\
A: I like natural scenery, but I especially like desert landscapes. It's quite difficult to see in Japan, though. What kind of scenery do you like? \\
B: Desert landscapes, indeed a world far from everyday life. I like cherry blossoms too, but I also find autumn foliage beautiful. \\
A: Both are uniquely Japanese seasonal features, aren't they? Makes you want to go mountain hiking. \\
B: I aspire to go mountain climbing. Though I have neither the stamina nor the gear. I'd like someone who enjoys that kind of thing to take me. \\
A: Mountain climbing certainly has its dangers. I once got worried I was lost while hiking alone, and it was truly terrifying. \\
B: That's your own experience, huh? Scary. But someday I want to try solo mountain climbing. \\
A: Nature is captivating because it's mysterious. Thank you for the enjoyable conversation. \\

\medskip
[Summary]\\ 
Is an indoor person and enjoys gaming as a hobby. Also likes walking, aiming to be conscious of posture and improve health. Mentioned enjoying walks in places with cherry trees, and likes natural scenery and autumn foliage. Aspires to go mountain climbing and is interested in the experience of solo mountain climbing.\\

\medskip
{\textbf{--Example2--}}\\\leavevmode
[Dialogue History]\\ 
A: Thank you for using our service today. Ma'am/Sir, are you here for a travel consultation today? \\
B: That's right. Yes, please. \\
A: My pleasure. Ma'am/Sir, where are you planning to travel? \\
B: Yes. Well, I've only vaguely decided, but, um, it's for a couple, a couple in their 50s, and since it's autumn, I was vaguely thinking of going towards the Hakone area. \\
A: I see, I see. That sounds like it will be a wonderful trip. \\
B: Haha, thank you. \\
A: Yes, so then, if I understand correctly, your request is for a trip for a couple, to Hakone, is that right? \\
B: Yes, uh, yes, ah, that's correct. Yes. \\
A: Understood. Then, I will first look into the Hakone area for you, so please wait a moment. \\
B: Yes, ah, please do. \\
A: Yes. So, Ma'am/Sir, regarding Hakone, is there anything specific you'd like to do, or perhaps a particular place you already want to visit? \\
B: Yes. \\
A: Or if you have a general idea of what you'd like to do, I can look into it for you. \\
B: Ah, yes. Um, well, there are a few places \texttt{<>}, but \\
A: I see. \\
B: Um, I don't know the exact names, but \\
A: I see. \\
B: I think there was something like a music box, ah, a music box museum or something like that \texttt{<>}, \\
A: I see, ah, I see. Ah, yes. \\
B: I'd like to do something cultural like that, yes. \\
A: I see, ah, I understand. Then, Ma'am/Sir, shall we look into that music box museum first? \\
B: Yes, ah, yes, please. \\
A: Please wait a moment. \\
B: Yes. \\
A: Yes, thank you for waiting. Regarding the music box museum, a place I can guide you to is, \\
B: Yes, yes. \\
A: Please wait a moment. This place, it's called Annoie, are you familiar with it, Ma'am/Sir? \\
B: Annoie? \\
A: Yes. \\
B: Um, how do you write ``An''? \\
A: It's written in Katakana as ``An''. \\
B: Ah, I see. ``An'' no, hmm, hmm, yes, ah, I didn't know that one. Yes. \\
A: Annoie, \texttt{<>}\\

\medskip
[Summary]\\ 
A couple in their 50s is planning a trip and is considering traveling to the Hakone area. They are thinking of visiting a music box museum.\\

\medskip
{\textbf{--Let's begin!--}}\\\leavevmode
[Dialogue History] \\ 

\textcolor{red!70!black}{\textbf{\{short\_dialogue\}}}\\

\medskip
[Summary]\\ 

\end{tcolorbox}

\begin{tcolorbox}[
  colback=gray!15,
  colframe=gray!70,
  coltitle=white,
  colbacktitle=gray,
  title=\texttt{Prompt for Generating Dialogue Summaries in Tabidachi Corpus}, 
  fonttitle=\bfseries\sffamily,
  fontupper=\footnotesize,
  boxrule=0.4mm,
  arc=2mm,
  left=2mm,
  right=2mm,
  top=1mm,
  bottom=1mm,
  sharp corners=south,
  enhanced jigsaw,
  breakable
]

\leavevmode Based on the following content, summarize B's hobbies and experiences regarding tourist destinations in a single sentence.\\ 
\leavevmode Please generate a summary sentence that includes as much information as possible.\\ 

\medskip
\leavevmode[Source Text for Summarization]\\\smallskip 
\textcolor{red!70!black}{\textbf{\{all\_short\_dialogue\_summary\}}}\\

\medskip
\leavevmode[Summary Sentence]\\ 

\end{tcolorbox}

\begin{tcolorbox}[
  colback=gray!15,
  colframe=gray!70,
  coltitle=white,
  colbacktitle=gray,
  title=\texttt{Prompt for Generating Dialogue Summaries in ChatRec}, 
  fonttitle=\bfseries\sffamily,
  fontupper=\footnotesize,
  boxrule=0.4mm,
  arc=2mm,
  left=2mm,
  right=2mm,
  top=1mm,
  bottom=1mm,
  sharp corners=south,
  enhanced jigsaw,
  breakable
]
\leavevmode You are a high-performance analytical assistant. Analyze the following dialogue history to extract and concisely summarize user \textcolor{red!70!black}{\textbf{\{user\}}}'s preferences, experiences, and hobbies.\\ 
\leavevmode 1. Task: Extract user \textcolor{red!70!black}{\textbf{\{user\}}}'s preferences, experiences, and hobbies from the dialogue history.\\ 
\leavevmode 2. Output format: Output as a sentence/text.\\ 
\leavevmode 3. Information to extract:\\ 
\leavevmode - Favorite activities (sports, arts, music appreciation, etc.)\\ 
\leavevmode - Favorite places or places they want to visit\\ 
\leavevmode - Favorite food/drinks\\ 
\leavevmode - Favorite entertainment (movies, TV, games, music, etc.)\\ 
\leavevmode - Things they are interested in\\ 
\leavevmode - Items they collect or collections\\ 
\leavevmode - Enjoyable activities they do regularly\\ 
\leavevmode 4. Information NOT to extract:\\ 
\leavevmode - Personality traits (kind, meticulous, etc.)\\ 
\leavevmode - Communication style\\ 
\leavevmode - Values or beliefs\\ 
\leavevmode - Characteristics of interpersonal relationships\\ 
\leavevmode - Patterns of emotional expression\\ 
\leavevmode - Characteristics of thought processes\\ 

\medskip
{\textbf{--Example--}}\\
\leavevmode[Dialogue History]\\ 
A: Nice to meet you.\\ 
B: Nice to meet you too. Did you go out during Golden Week this year?\\ 
A: Unfortunately, I couldn't go out because I was in an emergency declaration area. Did you go anywhere?\\ 
B: I also mostly stayed home, just went out for a bit of shopping. I feel like last year's Golden Week was similar.\\ 
A: That's right. It's been difficult to go out since COVID started, hasn't it?\\ 
B: Around the end of last year, I felt like it might be over by next year, but it's not at all.\\ 
A: I guess it will stay like this for a while longer. What do you want to do once it's over?\\ 
B: I want to travel to my heart's content, and visit museums and art galleries.\\ 
A: Traveling sounds nice. Domestic or overseas?\\ 
B: First, I want to travel domestically. I feel like relaxing in a place with clean air and beautiful greenery.\\ 
A: That sounds lovely. A place rich in nature would be very refreshing.\\ 
B: Yes, it's mentally tough when self-restraint continues. Are there any places you'd like to go after COVID?\\ 
A: Since the self-restraint has been long, I want to go to theme parks like Disneyland or music festivals and have fun.\\ 
B: Disneyland sounds great too. Come to think of it, I haven't been there at all since last year.\\ 
A: It was open, but it was hard to get tickets. I can't wait for the day we can go normally again.\\ 
B: You kind of feel restless if you don't go to Disneyland regularly, right? My kids really want to go.\\ 
A: Your children like Disney too, huh? The Beauty and the Beast attraction is new as well, so it should be fun again.\\ 
B: That's right, I hope it ends as soon as possible. Theme parks inevitably get crowded, don't they?\\ 
A: True. It might be a while longer, but I want to hang in there until COVID is over.\\ 
B: I hope vaccination progresses quickly. My mother finally had her first shot.\\ 
A: Oh, I see. I'm glad she was able to get an appointment and the shot safely. I hope we can get vaccinated soon too.\\ 

\medskip
\leavevmode[Summary]\\ 
Has been mostly staying home during the COVID-19 pandemic, only going out for essential shopping, and hopes to travel and visit museums/art galleries once the pandemic subsides. Prioritizes domestic travel and wants to relax in a nature-rich place with clean air and greenery. Has children who also like Disneyland and has a habit of visiting Disneyland regularly, but has not been able to go for over a year due to the pandemic. Has an elderly mother in the family who has already completed her first dose of the COVID-19 vaccine.\\ 

\medskip
{\textbf{--Let's begin!--}}\\
\leavevmode[Dialogue History]\\ 
\textcolor{red!70!black}{\textbf{\{dialogue\}}}\\

\medskip
\leavevmode[Summary of User \textcolor{red!70!black}{\textbf{\{user\}}}]\\ 
\medskip

\end{tcolorbox}
\subsection{Item Recommendation Information Prompt}
\label{appendix:rec_prompt}

\begin{tcolorbox}[
  colback=gray!15,
  colframe=gray!70,
  coltitle=white,
  colbacktitle=gray,
  title=\texttt{Prompt for Generating Item Recommendation Information in Tabidachi Corpus}, 
  fonttitle=\bfseries\sffamily,
  fontupper=\footnotesize,
  boxrule=0.4mm,
  arc=2mm,
  left=2mm,
  right=2mm,
  top=1mm,
  bottom=1mm,
  sharp corners=south,
  enhanced jigsaw,
  breakable
]
\leavevmode{\textbf{[TASK]}} \\ 
\leavevmode Generate a tourist destination recommendation information based on the tourist destination information. The recommendation information should be a single sentence.\\ 

\medskip
\leavevmode{\textbf{==Example1==}}\\ 
\leavevmode[Tourist Destination Information]\\ 
\leavevmode Located in the southeastern part of Sapporo city, this is an all-weather indoor dome. It is the home stadium for Hokkaido Consadole Sapporo and Hokkaido Nippon-Ham Fighters, hosting soccer and baseball games, as well as various other events like sports and concerts. There are shops for enjoying shopping and dining, and an observation deck with a great view. On non-event days, dome tours offering a behind-the-scenes look at Sapporo Dome are also held. Duration: 90 minutes or more. Recommended for babies, kids, women, winter, and rainy days.\\ 

\medskip
\leavevmode[Tourist Destination Recommendation Information]\\ 
\leavevmode This all-weather indoor dome is the home stadium for Hokkaido Consadole Sapporo and Hokkaido Nippon-Ham Fighters, hosting not only soccer and baseball games but also various sports and concert events, along with offering shopping and dining options, and an observation deck with a great view, making it a recommended tourist spot for those with children or those who want to sightsee even on rainy days.\\ 

\medskip
\leavevmode{\textbf{==Let's begin!==}}\\ 
\leavevmode[Tourist Destination Information]\\ 
\textcolor{red!70!black}{\textbf{\{rec\_info\}}}\\

\leavevmode[Tourist Destination Recommendation Information]\\ 

\medskip

\end{tcolorbox}

\begin{tcolorbox}[
  colback=gray!15,
  colframe=gray!70,
  coltitle=white,
  colbacktitle=gray,
  title=\texttt{Prompt for Generating Item Recommendation Information in ChatRec}, 
  fonttitle=\bfseries\sffamily,
  fontupper=\footnotesize,
  boxrule=0.4mm,
  arc=2mm,
  left=2mm,
  right=2mm,
  top=1mm,
  bottom=1mm,
  sharp corners=south,
  enhanced jigsaw,
  breakable
]
\leavevmode You are a professional tourist destination recommender.\\ 
\leavevmode Generate a tourist destination recommendation information based on the tourist destination information. The recommendation information should be a single sentence.\\ 

\medskip
\leavevmode\textbf{--Example1--}\\
\leavevmode[Tourist Destination Information]\\ 
\leavevmode Located in the southeastern part of Sapporo city, this is an all-weather indoor dome. It is the home stadium for Hokkaido Consadole Sapporo and Hokkaido Nippon-Ham Fighters, hosting soccer and baseball games, as well as various other events like sports and concerts. There are shops for enjoying shopping and dining, and an observation deck with a great view. On non-event days, dome tours offering a behind-the-scenes look at Sapporo Dome are also held. Duration: 90 minutes or more. Recommended for babies, kids, women, winter, and rainy days.\\ 

\medskip
\leavevmode[Tourist Destination Recommendation information]\\ 
\leavevmode This all-weather indoor dome is the home stadium for Hokkaido Consadole Sapporo and Hokkaido Nippon-Ham Fighters, hosting not only soccer and baseball games but also various sports and concert events, along with offering shopping and dining options, and an observation deck with a great view, making it a recommended tourist spot for those with children or those who want to sightsee even on rainy days.\\ 

\medskip
\leavevmode\textbf{--Let's begin!--}\\
\leavevmode[Tourist Destination Information]\\ 
\textcolor{red!70!black}{\textbf{\{rec\_info\}}}\\

\medskip
\leavevmode[Tourist Destination Recommendation Information] 

\end{tcolorbox}

\newpage
\twocolumn

\section{Implementation and Hyperparameter Details} 
\label{appendix:hyper-parameter}
Our framework was implemented in Python 3.10.12.
We used the following Python package versions to conduct all experiments:

\begin{itemize}
    \item \textbf{PyTorch} (version 2.4.1)
    \item \textbf{Hugging Face Transformers} (version 4.46.2)
    \item \textbf{Hugging Face Tokenizers} (version 0.20.3)
    \item \textbf{SacreBLEU} (version 2.5.1)
    \item \textbf{rouge-score} (version 0.1.2)
    \item \textbf{Fugashi} (version 1.4.0) with \textbf{MeCab} 
    \item \textbf{Optuna} (version 4.1.0)
    \item \textbf{Hugging Face Datasets} (version 3.1.0)
    \item \textbf{Hugging Face TRL (Transformer Reinforcement Learning)} (version 0.12.1)
\end{itemize}

Tables \ref{tab:tabidachi-dpo-hyper} and \ref{tab:chatrec-dpo-hyper} list the hyperparameters used for fine-tuning (i) the dialogue summary generation model and (ii) the item recommendation information generation model with DPO. The hyperparameter optimization was performed using Optuna ~\citep{optuna}. Subsequently, for both the dialogue summary generation model and the item recommendation information generation model, we selected the hyperparameters that yielded the best recommendation performance on the validation set. Each model was then trained five times using these selected hyperparameters, and the average results from these five trained models were used as the final results in this study.

We ran all experiments primarily on four Nvidia A100 80GB GPUs.
For the Llama-3.1-swallow-8B model, training was conducted for a single epoch. Each such training run required approximately 24 hours using all four GPUs.
For the DeBERTa model, we performed hyperparameter tuning by training it for 1, 4, and 10 epochs. A single epoch of DeBERTa training took approximately 4 hours on the four-GPU setup.

\paragraph{Licensing.}
\textbf{Tabidachi Corpus} is released under the Creative Commons Attribution 4.0 (CC BY 4.0) license.
\textbf{ChatRec} (dataset + baseline code) is distributed under the permissive MIT License.
The pretrained language models employed in our pipeline follow different terms: %
\textbf{Llama-3.1-Swallow-8B} is governed by the Meta Llama 3.1 Community License together with the Gemma Terms of Use, which allow research and commercial use subject to the stated usage restrictions; %
\textbf{deberta-v3-japanese-large} is released under Creative Commons Attribution-ShareAlike 4.0 (CC BY-SA 4.0).

\begin{table*}[ht]
  \centering
  \small
  \begin{tabular}{@{}p{0.36\linewidth}p{0.28\linewidth}p{0.28\linewidth}@{}}
    \toprule
    \textbf{Parameter} & \textbf{Summary Model (DPO)} & \textbf{Recommendation Model (DPO)}\\
    \midrule
    learning\_rate & $1.1593\times10^{-7}$ & $8.7340\times10^{-6}$\\
    per\_device\_train\_batch\_size & 12 & 16\\
    num\_train\_epochs & 1 & 1\\
    optimizer & AdamW ($\beta_{1}=0.9$, $\beta_{2}=0.999$, $\epsilon=10^{-8}$, weight\_decay$=0$) &
                AdamW ($\beta_{1}=0.9$, $\beta_{2}=0.999$, $\epsilon=10^{-8}$, weight\_decay$=0$)\\
    max\_grad\_norm & 1.0 & 1.0\\
    gradient\_checkpointing & True & True\\
    bf16 & True & True\\
    disable\_dropout & True & True\\
    \midrule
    \multicolumn{3}{@{}l}{\textbf{DPO‐Specific Parameter}}\\
    \midrule
    $\beta$ & 0.1768 & 0.06109\\
    \bottomrule
  \end{tabular}
  \caption{Hyperparameters for the summary generation model and item recommendation information generation model, fine-tuned using DPO on Tabidachi Corpus.}
  \label{tab:tabidachi-dpo-hyper}
\end{table*}

\begin{table*}[ht]
  \centering
  \small
  \begin{tabular}{@{}p{0.36\linewidth}p{0.28\linewidth}p{0.28\linewidth}@{}}
    \toprule
    \textbf{Parameter} & \textbf{Summary Model (DPO)} & \textbf{Recommendation Model (DPO)}\\
    \midrule
    learning\_rate & $6.4087\times10^{-7}$ & $1.7718\times10^{-7}$\\
    per\_device\_train\_batch\_size & 8 & 8\\
    num\_train\_epochs & 1 & 1\\
    optimizer & AdamW ($\beta_{1}=0.9$, $\beta_{2}=0.999$, $\epsilon=10^{-8}$, weight\_decay$=0$) &
                AdamW ($\beta_{1}=0.9$, $\beta_{2}=0.999$, $\epsilon=10^{-8}$, weight\_decay$=0$)\\
    max\_grad\_norm & 1.0 & 1.0\\
    gradient\_checkpointing & True & True\\
    bf16 & True & True\\
    disable\_dropout & True & True\\
    \midrule
    \multicolumn{3}{@{}l}{\textbf{DPO‐specific Parameter}}\\
    \midrule
    $\beta$ & 0.1253 & 0.03949\\
    \bottomrule
  \end{tabular}
  \caption{Hyperparameters for the summary generation model and item recommendation information generation model, fine-tuned using DPO on the ChatRec dataset.}
  \label{tab:chatrec-dpo-hyper}
\end{table*}

\section{Human Evaluation Details}
\begin{figure}[ht]
    \centering
    \includegraphics[width=\columnwidth]{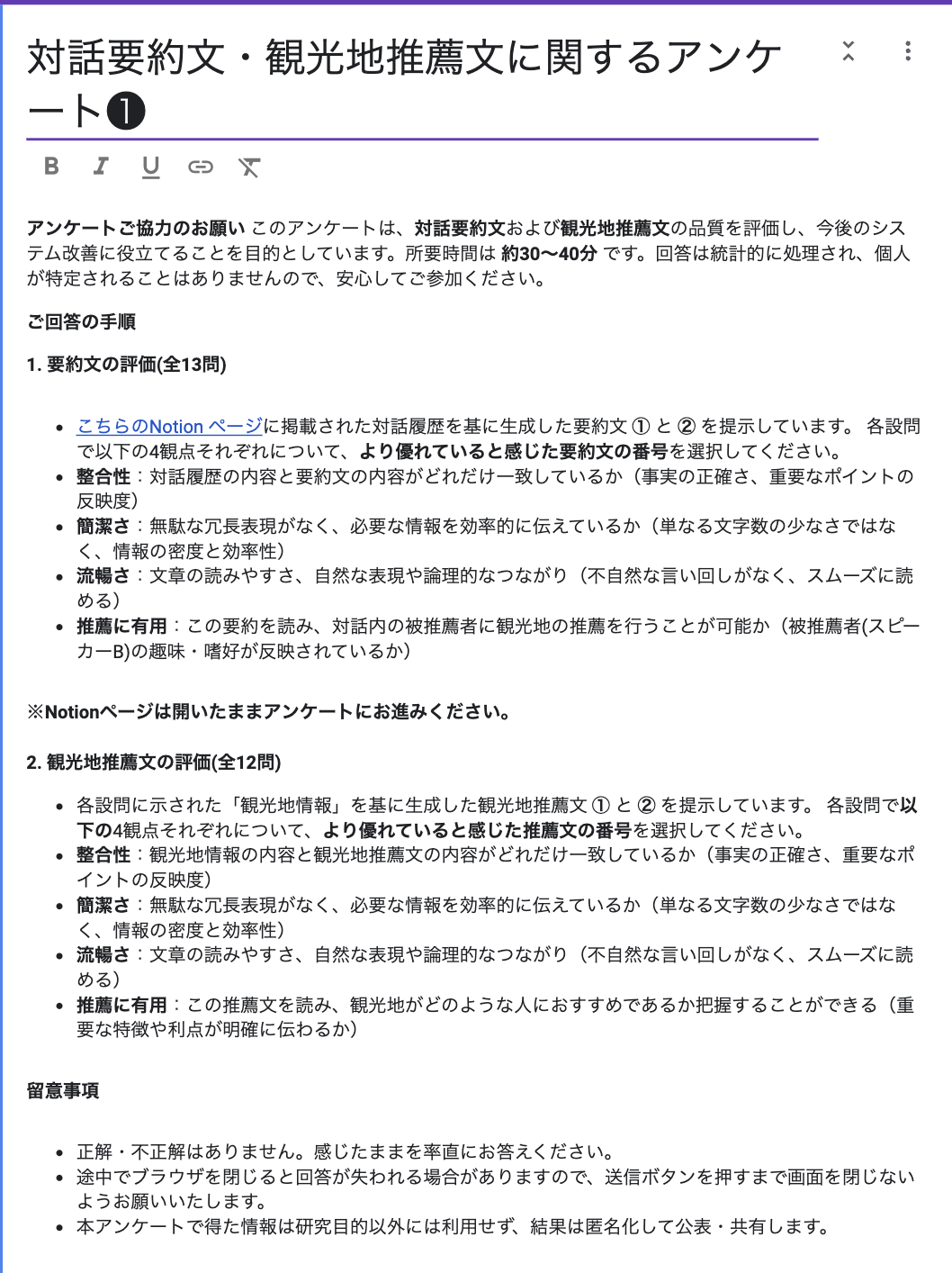}
    \caption{Requests to Crowd Workers}
    \label{human_eval1}
\end{figure}

\begin{figure}[ht]
    \centering
    \includegraphics[width=\columnwidth]{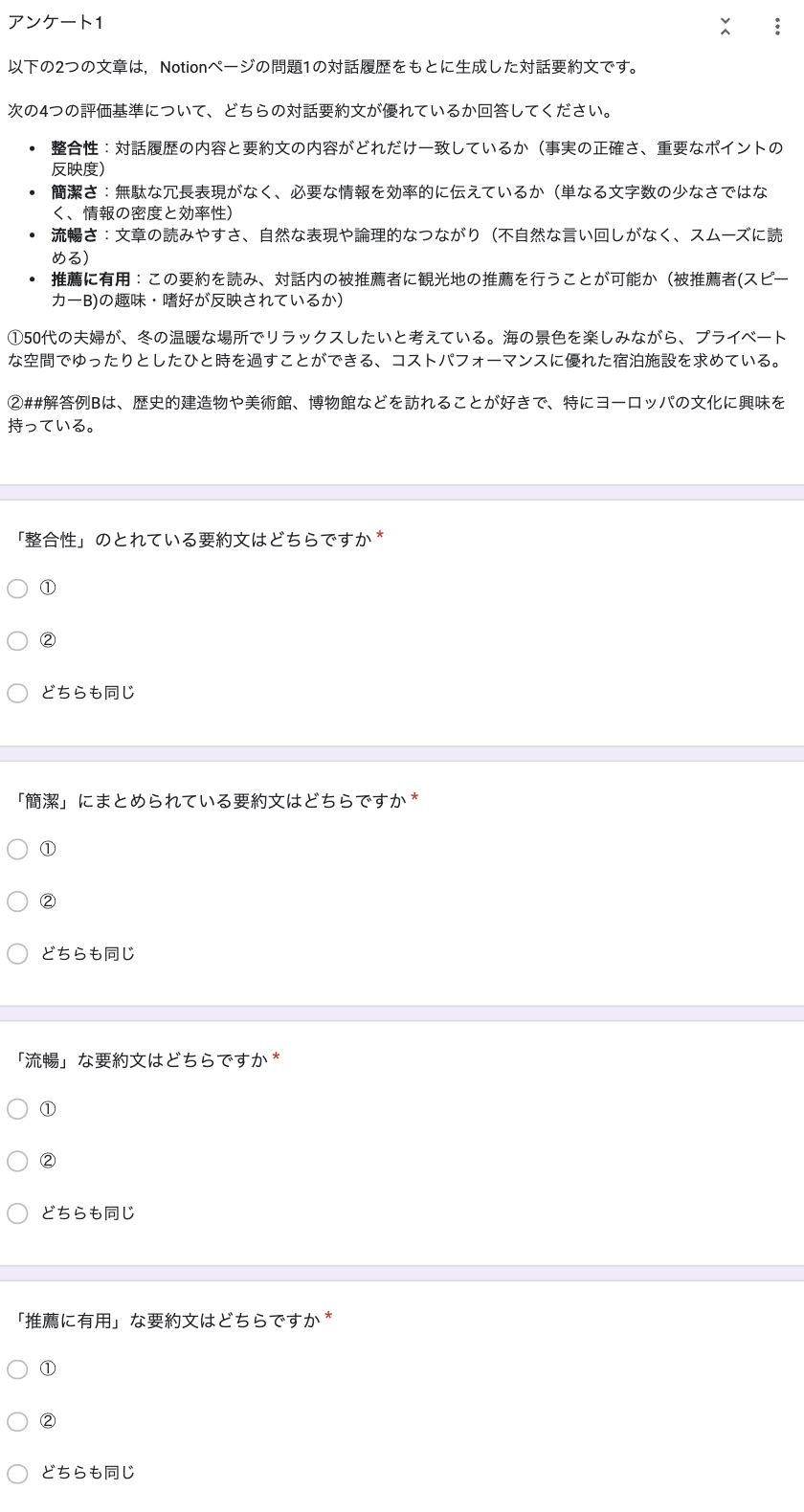}
    \caption{Crowdworker response screen}
    \label{human_eval2}
\end{figure}

\begin{figure}[ht]
    \centering
    \includegraphics[width=\columnwidth]{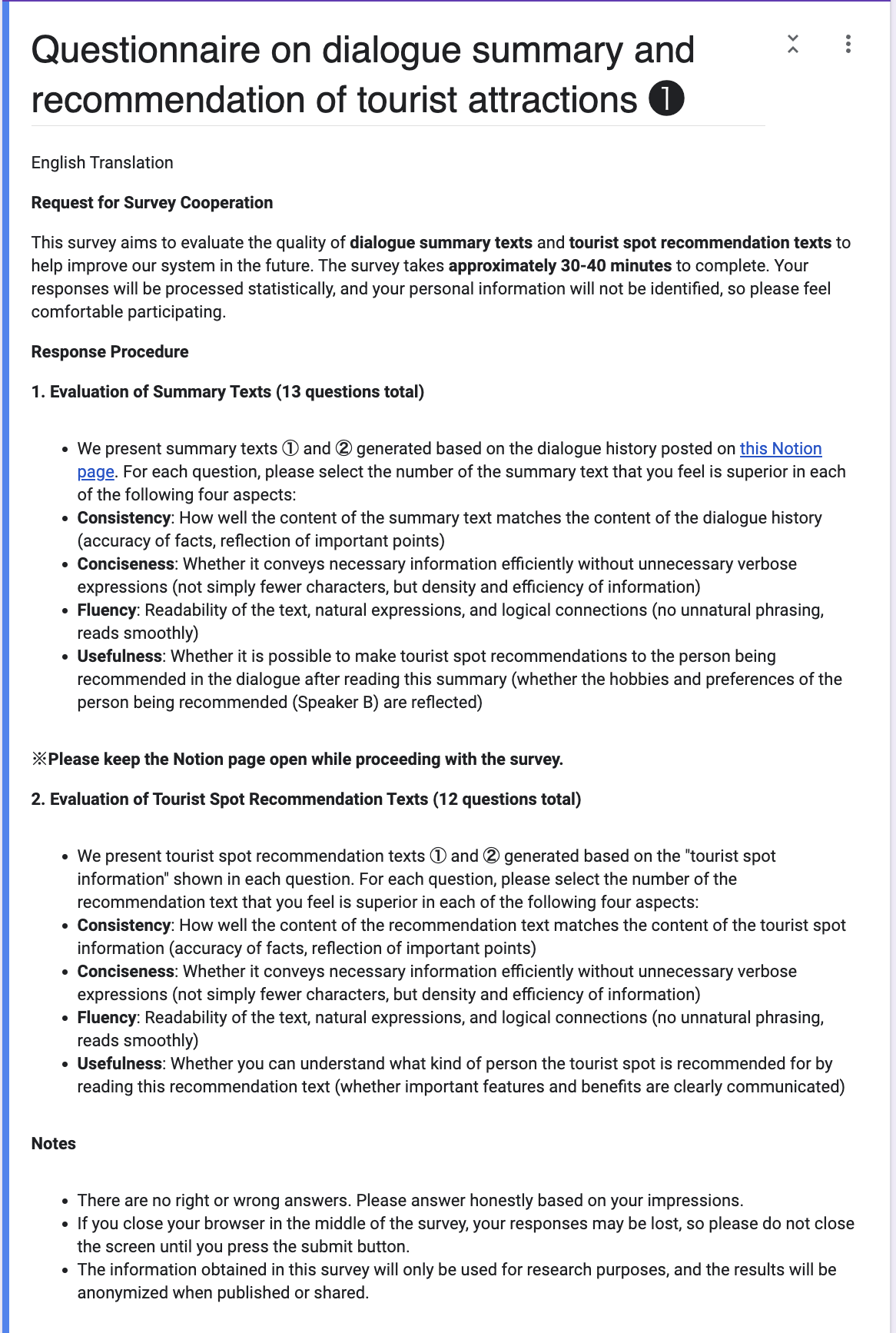}
    \caption{Requests to Crowd Workers(English Version, translated by Author)}
    \label{human_eval3}
\end{figure}

\begin{figure}[ht]
    \centering
    \includegraphics[width=\columnwidth]{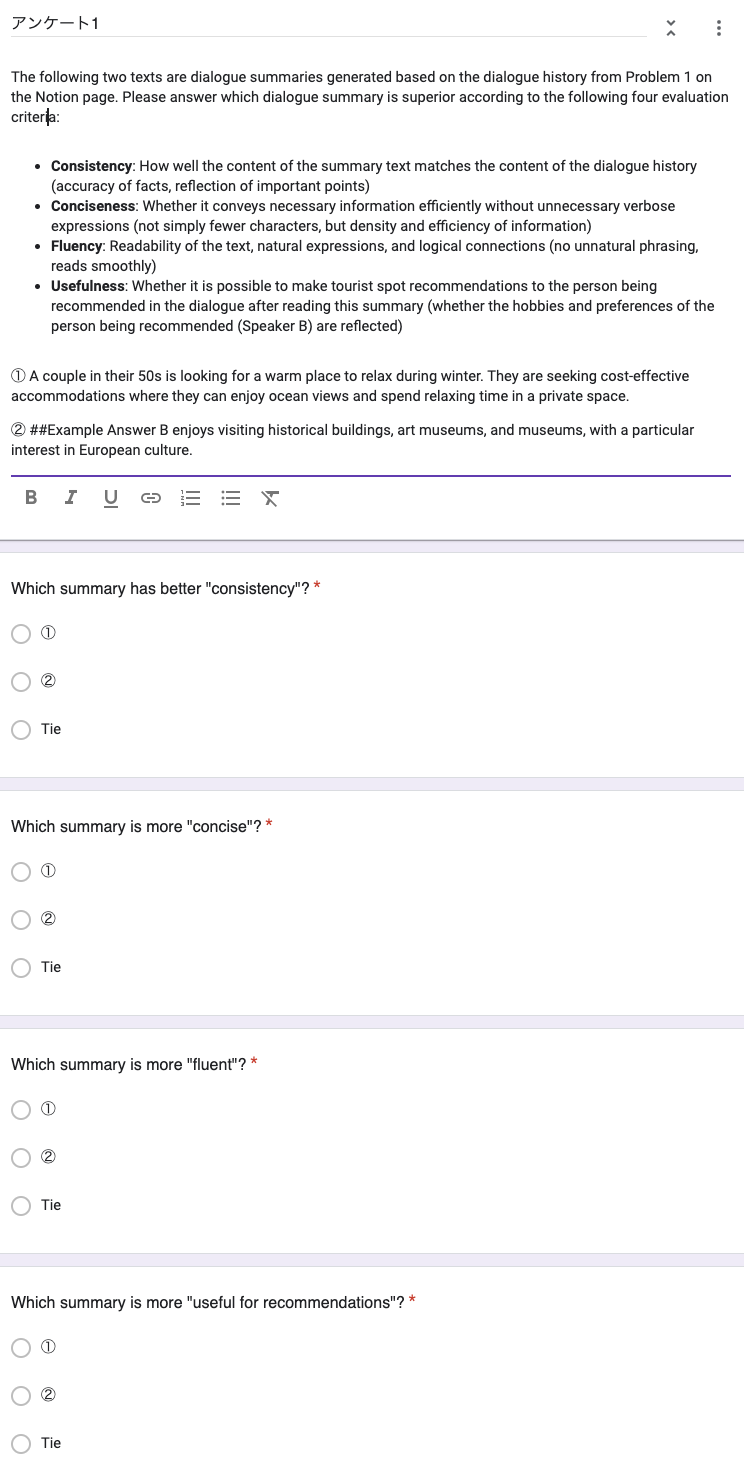}
    \caption{Crowdworker response screen (English Version, translated by Author)}
    \label{human_eval4}
\end{figure}

Participants were compensated 700 JPY per questionnaire. This rate was determined considering that each questionnaire required approximately 30 to 40 minutes to complete, and by taking into account the minimum wage in Japan to ensure fair compensation for their time.Furthermore, screenshots of the actual questionnaires used are provided in Table\ref{human_eval1}, \ref{human_eval2}, \ref{human_eval3}, \ref{human_eval4}.

\section{Case Study}
\label{appendix:case-study}
Table \ref{abstract-table} shows examples of dialogue summaries and item recommendation information generated by our proposed method and SumRec. The dialogue summaries reflect customer preferences, such as ``wishing to stay on a remote island'' or ``being interested in a tour along the coast in an ox-drawn cart and dishes made with Agu pork.'' Furthermore, examples of item recommendation information include information about what kind of user an item is suitable for, such as ``perfect for a date with a loved one or an anniversary dinner'' or ``you can enjoy delicious food at a reasonable price, so it's easy on your wallet.''

Table \ref{tab:rec-result} presents an example where information not present in the original item description is included in the generated item recommendation information. In this instance, the phrase ``the interior is stylish and has a calm atmosphere'' is included in the item recommendation information, despite this detail not being part of the item description. This can make it difficult to accurately predict scores in some cases. Future work needs to focus on improving the system to generate item recommendation information that are strictly based on factual information.

\begin{table*}[ht]
\small
\renewcommand{\arraystretch}{1.15}      
\setlength{\tabcolsep}{3pt}             
\centering

\begin{tabular*}{\textwidth}{
        @{}                               
        p{.25\textwidth}                
        @{\hspace{3pt}}                   
        p{\dimexpr.75\textwidth-3pt}    
        @{}
      }
\hline
\textbf{Dialogue History} &
\begin{minipage}[t]{\linewidth}\raggedright\small
Operator: Thank you for using our service today.\par
Customer: My pleasure.\par
Operator: Thank you. Ma'am/Sir, are you here for a travel consultation today?\par
... (omitted) ...\par
Customer: Hmm. If possible, I'd like to visit a remote island.\par
... (omitted) ...\par
Operator: Yes, that's right... Ma'am/Sir, earlier you mentioned you like Agu pork...\par
Customer: Yes. Uh-huh. Motobu beef, yes.\par
... (omitted) ...\par
Customer: Ah, is that your top recommendation?\par
Operator: Hmm, there might be better places—\texttt{<>} well, ah.\par
Customer: \texttt{<>} For now, around there, yes.
\end{minipage}\\
\hline

\textbf{Item description} &
\begin{minipage}[t]{\linewidth}\raggedright\small
Enjoy reasonably priced dishes made with carefully selected ingredients, including A5 and A4 grade Motobu beef, Agu pork, Yanbaru young chicken, and exceptionally fresh seafood chosen by the head chef. Duration: 30-60 minutes / English menu available / 3000-5000 yen (dinner) / Recommended for business entertainment.
\end{minipage}\\
\hline

\textbf{SumRec Dialogue Summary} &
\begin{minipage}[t]{\linewidth}\raggedright\small
A woman planning a trip to Okinawa wants to gaze at the sea in beach sandals and ride an ox-drawn cart. She is also looking forward to Agu pork dishes.
\end{minipage}\\
\hline

\textbf{SumRec Item Recommendation Information} &
\begin{minipage}[t]{\linewidth}\raggedright\small
This is a restaurant located in Motobu Town, Okinawa Prefecture. You can enjoy dishes made with A-grade beef and fresh seafood. It is also recommended for business entertainment.
\end{minipage}\\
\hline

\textbf{Proposed Method Dialogue Summary} &
\begin{minipage}[t]{\linewidth}\raggedright\small
Ms. B (a single woman in her 50s) planning a trip to Okinawa wants to enjoy a relaxing time by the sea and also \textbf{wishing to stay on a remote island}. She is also \textbf{being interested in a tour along the coast in an ox-drawn cart and dishes made with Agu pork}. Ms. A is introducing stores and spots that Ms. B might like. 
\end{minipage}\\
\hline

\textbf{Proposed Method Item Recommendation Information} &
\begin{minipage}[t]{\linewidth}\raggedright\small
Located in the resort area of northern Okinawa main island, you can savor dishes made with A-rank Kuroge Wagyu steak and abundant local seafood. The restaurant has a calm atmosphere, \textbf{perfect for a date with a loved one or an anniversary dinner}. Private rooms are also available for a relaxing time with family or friends. \textbf{You can enjoy delicious food at a reasonable price, so it's easy on your wallet}. 
\end{minipage}\\
\hline
\end{tabular*}
\caption{Example of output sentences before and after DPO}
\label{abstract-table}
\end{table*}

\begin{table*}[ht]
  \small
  \centering
  \begin{tabular}{p{0.25\textwidth}|p{0.75\textwidth}}
    \hline
    \textbf{Item Information} &
    KiKiYOKOCHO is a new concept zone that gathers items to tickle women's sensibilities by mixing beauty, food, and miscellaneous goods. The concept is ``try, find, enjoy.'' For those who want to compare and try things they are interested in to find what matches their personal preferences. A collection of shops that fulfills such selfish desires. It's packed with unprecedented enjoyment. Duration: around 30-60 minutes. English pamphlets available. \\
    \hline
    \textbf{Item Recommendation Information by Proposed Method} &
    This is a shopping mall targeted at women, featuring stores from various genres such as beauty, gourmet, fashion, and interior design. \textbf{The interior is stylish and has a calm atmosphere}, allowing you to enjoy shopping at a leisurely pace. Additionally, English signboards are available, so foreign visitors can also use it with peace of mind. The shop staff are also kind and helpful, so even first-time visitors can visit casually. \\ 
    \hline
  \end{tabular}
  \caption{Example of an item recommendation information containing incorrect information}
  \label{tab:rec-result}
\end{table*}

Table \ref{tab:winner-loser} shows an example of the positive (winner) and negative (loser) samples used for DPO training. This process facilitates the training of text generation models capable of outputting texts that enable the score predictor to accurately interpret user preferences, experiences, and item descriptions.

\begin{table*}[ht]
  \small
  \centering
  \begin{tabular}{p{0.25\textwidth}|p{0.75\textwidth}}
    \hline
    \textbf{Winner} &
    Rows of old buildings create an atmosphere of traditional Japan. There is also a spacious park where families with children can enjoy themselves with peace of mind. Within the park, there is an exhibition hall where visitors can learn about the region's traditions and culture, making it an attractive spot especially for families. In particular, the area is cool and pleasant in summer, making it highly recommended for family visits. \\
    \hline
    \textbf{Loser} &
    Rows of old buildings stand, and various exhibitions are held. There is also a large park where families with children can play safely. Visitors can also enjoy light hiking and experience nature. In summer, it is cool and an ideal place for children to have fun. \\
    \hline
  \end{tabular}
  \caption{Examples of positive (Winner) and negative (Loser) Item Recommendation Information used in DPO}
  \label{tab:winner-loser}
\end{table*}

\end{document}